\documentclass{elsart}

\usepackage{graphicx}

\usepackage{amssymb}

\begin{document}
\journal{Solid State Communications}

\begin{frontmatter}



\title{Quantum dots as scatterers in electronic transport : interference and correlations}


\author{Piotr Stefa\'{n}ski}

\address{Institute of Molecular Physics of the Polish Academy of Sciences,\\
Smoluchowskiego 17, 60-179 Pozna\'{n}, Poland}

\begin{abstract}
Conductance  through a system consisting of a wire with
side-attached quantum dots is calculated. Such geometry of the
device allows to study the coexistence of quantum interference,
electron correlations and their influence on conductance. We
underline the differences between "classical" Fano resonance in
which the resonant channel is of single-particle nature and
"many-body" Fano resonance with the resonant channel formed by
Kondo effect. The influence of electron-electron interactions on
the Fano resonance shape is also analyzed.
\end{abstract}

\end{frontmatter}

\section{\label{sec:1}Introduction}
A model consisting of a quantum dot (QD) attached to a metallic
quantum wire is conceptually analogous to the Fano model
\cite{fano} consisting of a continuous spectrum and a discrete
level. Such geometry is also in close relation to the behavior of
magnetic impurities embedded in the host metal, where the
conduction electrons are scattered resonantly on impurities while
propagating through the metal. This arrangement can be opposed to
the standard geometry, where a quantum dot is connected in series
with leads and resonantly enhances the conductance. A model of a
quantum dot side-coupled to the quantum wire is also applicable
for the simplest description of a magnetic atom deposited on the
metallic surface. In such a "mapping" the QD in Kondo regime
mimics the adatom with nonzero magnetic moment and the wire (when
assumed to be multichannel) serves as an analog of the metallic
surface host. Thus, the model considered by us can provide
information applicable to various nanostructures.

 Fano effect, described in 1961, has been observed in
many physical systems: in nuclear scattering \cite{adair},
scattering from donor impurities in an electron waveguide
\cite{chu}, optical absorbtion \cite{maschke}, electronic
transport in heterojunctions \cite{ting}. Recently it is seen in
new light due to the rapid progress in nanotechnology. Apart from
'classical' Fano effect, where the resonant channel is provided by
discrete single-particle level, a new type emerged, where the
resonant channel is formed by many-body effects. This additional
resonant channel is formed by the electron-electron interactions
resulting in the sharp Abrikosov-Suhl (Kondo) resonance in the
vicinity of Fermi surface. This kind of Fano resonance has been
observed recently in scanning tunnelling microscope experiments
for single atoms placed on metallic surface \cite{li,madhavan} and
also in lateral quantum dots \cite{gold}. There were theoretical
attempts to describe these phenomena in \cite{ujsaghy} and in
\cite{hofstetter}, respectively.

A single quantum dot modelled by Anderson Hamiltonian and
side-attached to a perfect wire was investigated in resonant
regime by slave boson mean field approach \cite{kang}, exact
diagonalization \cite{torio}, interpolative perturbative scheme
(IPS) \cite{aligia} and $X$- boson treatment \cite{franco}. For
this arrangement we concentrate on the crossover between
"classical" and "many-body" Fano resonances.

A system of two quantum dots placed in arms of Aharonov-Bohm ring
has been investigated recently for non-interacting electrons
\cite{kubala}. In our model we study the influence of the strong
electron correlations within the dots on transport. Repulsive
Coulomb interactions in the nanoscale devices are defined by the
charging energy $\sim e^{2}/2C$, where $C$ is the capacitance of
the device and $e$- electronic charge. They cause the most
celebrated phenomena as Coulomb blockade and resonant (Kondo)
electron transport \cite{gold}.
\section{\label{sec:2}Model and calculations}
 Hamiltonian of the nanodevice consisting of metallic wire and quantum dots attached parallel to
 it, is taken in the form:
\begin{eqnarray}
\label{for1} \nonumber
H=\sum_{k,\sigma}\epsilon_{k}c_{k,\sigma}^{+}c_{k,\sigma}
 +\sum_{\gamma=1,2}\sum_{\sigma}\epsilon_{\gamma}d_{\gamma,\sigma}^{+}d_{\gamma,\sigma}\\
+
\sum_{\gamma=1,2}U_{\gamma}n_{\gamma\uparrow}n_{\gamma\downarrow}+
 \sum_{\gamma=1,2\atop\sigma,k}t_{\gamma}\lbrack
 d_{\gamma\sigma}^{+}c_{k\sigma}+h.c.
 \rbrack
\end{eqnarray}
Creation operator $c_{k,\sigma}^{+}$ ($d_{\gamma,\sigma}^{+}$)
describes the conduction electron in the wire with momentum $k$,
energy $\epsilon_{k}$ and spin $\sigma$ ($\gamma$-th QD localized
electron with spin $\sigma$). $n_{\gamma,\sigma}$ is the number of
electrons with the spin $\sigma$ at the localized state $\gamma$
with the Coulomb repulsion $U_{\gamma}$. The last term describes
hopping between QDs and wire. We assume spin-only degeneracy of
the discrete QD's levels and the wire to be single-channel. The
subscript $\gamma$ represents the orbital quantum number, which
follows from quantization due to spatial confinement of the 2D
electron gas within the quantum dot (see for example \cite{kou}).
We underline that $\gamma$ is not conserved in the process of
hopping of electron between QD and causes an indirect interaction
between the dots.

To investigate transport through nanodevice, the wire has been
connected to external electrodes: $H_{con}=\sum_{\alpha=L,R\atop
k,k',\sigma}t_{\alpha}\lbrack
c_{k\sigma,\alpha}^{+}c_{k'\sigma}+h.c.\rbrack$;
$c_{k\sigma,\alpha}^{+}$ is the creation operator of the electron
with the spin $\sigma$ and energy $\epsilon_{k}$ in the electrode
$\alpha$. Spectral densities of electronic states in the wire
$\rho_{w0,\sigma}(\omega)$ and in the electrodes
$\rho_{el,\sigma}(\omega)$ have been assumed to have Lorentzian
shape with a halfwidth much larger than Kondo temperature of each
QD.

The numerical parameters of the equal Coulomb repulsion
$U_{\gamma}=$ 1 meV (for $\gamma=$1, 2) and tunneling strength
$\Gamma_{\gamma,\sigma}(\omega=\epsilon_{F})=2\pi
t_{\gamma}^{2}\rho_{w,\sigma}(\omega=\epsilon_{F})\equiv\Gamma_{\gamma
max,\sigma}=$ 0.28 meV have been chosen to meet the experimental
QD's data ($\epsilon_{F}=0$ defines the Fermi level in the wire).
A scheme of the investigated nanodevice is presented in the
$a$-inset of Fig.(\ref{fig2}).

Following Meir and Wingreen \cite{meir}, conductance can be
related to the spectral density of the central region (in the
present case - metallic quantum wire with quantum dots attached to
it) and its coupling $\Gamma_{L(R),\sigma}(\omega)=2\pi
t_{L(R)}^{2}\rho_{el,\sigma}(\omega)$ to left (L) and right (R)
electrode. For the symmetric coupling and in the zero limit of
drain-source voltage conductance can be written in the form:
 \begin{eqnarray}
 \label{for4}
 \mathcal{G}=\frac{2\pi e^{2}}{h}\sum_{\sigma}\int_{-\infty}^{\infty}\Gamma_{\sigma}(\epsilon)\left(-\frac{\partial f(\epsilon)}{\partial
 \epsilon}\right)\rho_{w,\sigma}(\epsilon)d\epsilon,
 \end{eqnarray}
where $f(\epsilon)$  is the Fermi distribution function and
$\Gamma_{\sigma}(\epsilon)=\Gamma_{L,\sigma}(\epsilon)\Gamma_{R,\sigma}(\epsilon)/(\Gamma_{L,\sigma}(\epsilon)
+\Gamma_{R,\sigma}(\epsilon))$. Spectral density of the
investigated nanodevice is calculated from the appropriate
retarded Green function: $\rho_{w,\sigma}(\omega)=-(1/\pi)\rm{Im}
G_{\sigma}(\omega+i\delta)$. At $T=0$, when $-\frac{\partial
f(\epsilon)}{\partial\epsilon} \to \delta(\epsilon)$ the formula
for conductance takes a simple form $\mathcal{G}=\frac{2\pi
e^{2}}{h}\sum_{\sigma}\Gamma_{\sigma}(0)\rho_{w,\sigma}(0)$.

To calculate conductance through the considered nanodevice the
Green function of the electronic wave propagating through the wire
should be calculated {\it in presence} of QDs attached to the
wire. We start from the Dyson equation (energy and spin dependence
of the Green functions and selfenergies is understood):
\begin{equation}
\label{dys}
 G=G_{0}+G_{0}\Sigma G,
\end{equation}
which can be written by iterating in the form of T-matrix:
\begin{eqnarray}
\label{dys1}
 G=G_{0}+G_{0}T G_{0},\; T=\Sigma /(1-\Sigma G_{0}).
\end{eqnarray}
$G_{0}$ is a free conduction electron Green function. The form of
selfenergy (and T-matrix) depends on the geometry of nanodevice
and approximation made.

Dyson equation describing scattering of the electronic wave on
both quantum dots:
\begin{eqnarray}
\label{2qddys} \nonumber G=G_{0}+G_{0} \Sigma_{1} G_{0}+G_{0}
\Sigma_{2} G_{0}+G_{0} \Sigma_{1} G_{0}\Sigma_{2} G_{0}\\ +G_{0}
\Sigma_{2} G_{0}\Sigma_{1} G_{0}+ ...
\end{eqnarray}
can be cast into the form Eq.(~\ref{dys}), when
$\Sigma=\Sigma_{1}+\Sigma_{2}$. The selfenergy $\Sigma_{\gamma}$,
($\gamma=1,2$) arises from scattering of the conduction electron
propagating through the wire on the $\gamma$-th quantum dot. Thus,
the total selfenergy $\Sigma$ contains infinite number of
scattering events in quantum dot $\gamma$ alone and multiple
scatterings involving both quantum dots. Utilizing the second part
of Eq.(\ref{dys1}) we can express $\Sigma_{1}$ and $\Sigma_{2}$ by
$T_{1}$ and $T_{2}$ and the selfenergy due to both quantum dots
is:
\begin{equation}
\label{sigtot}
\Sigma=\sum_{\gamma}\Sigma_{\gamma}=\frac{T_1}{(1+T_1G_0)}+\frac{T_2}{(1+T_2G_0)}.
\end{equation}
For non-interacting electrons, the expression for T-matrix due to
one QD attached to the wire obtained by equation of motion (EOM)
for conduction electron Green function is of the form
$T_{\gamma,\sigma}(\omega)=t_{\gamma}^{2}
G_{QD\gamma,\sigma}(\omega)$, where the Green function of the dot
$ G_{QD\gamma}(\omega)$ represents a simple single-particle level.
In the interacting case, it is replaced by dressed Green function
of the localized QD level, which has to be calculated in such
approximation that properly describes Kondo effect. In the present
work the Interpolative Perturbative Scheme (IPS) is utilized
\cite{yeyati,kajueter,potthoff}. It is based on the selfconsistent
second order perturbation (SOPT) in Coulomb repulsion U
\cite{YY,horv}. However, within the IPS the calculated selfenergy
due to strong electron correlations $\Sigma^{IPS}$ is taken in the
form, that has correct limits: when $U \to 0$, $\Sigma^{IPS} \to
\Sigma^{SOPT}$ and when $\Gamma \to 0$, $\Sigma^{IPS} \to
\Sigma^{atom}$. This allows to calculate the conductance both in
resonant regime and in Coulomb blockade regime. Moreover, by
replacing Hartree-Fock solution for the impurity level by an
effective field determined selfconsistently from the condition on
particle number, the fulfillment of the Friedel-Langreth sum rule
\cite{FL} is also obtained within this method. It is applicable
both for $T=0$  and finite temperatures.

If the expressions for $T_{1}$ and $T_{2}$-matrices  are known,
then the total selfenergy Eq.(\ref{sigtot}) is calculated and
finally total T-matrix by second part of Eq.(\ref{dys1}):
\begin{equation}
\label{totalT} T=\frac{T_{1}(1+T_{2}G_{0})+T_{2}(1+T_{1}G_{0})}
{(1+T_{1}G_{0})(1+T_{2}G_{0})-G_{0}[T_{1}(1+T_{2}G_{0})+T_{2}(1+T_{1}G_{0})]}.
\end{equation}
The extension of the above formalism on $\gamma=N$ QDs connected
to the wire in a star-like fashion is straightforward.
\section{One quantum dot attached to the wire}
Utilizing Eq.(\ref{dys1}) spectral density for the propagating
electron can can be written in the following form:
\begin{eqnarray}
\nonumber \rho_{w,\sigma}(\omega)=\rho_{w0,\sigma}(\omega)\times\
\label{ro1qd}\\\lbrace1+ImG_{0,\sigma}(\omega)[ImT_{\sigma}(\omega)
(q_{\sigma}^2-1)-2q_{\sigma}ReT_{\sigma}(\omega)]\rbrace,
\end{eqnarray}
where we have denoted:
\begin{equation}
\label{Fanoq}
 q_{\sigma}=-\frac{ReG_{0,\sigma}(\omega + i\delta)}{ImG_{0,\sigma}(\omega +
i\delta)}=\frac{\Delta_{\sigma}}{\Gamma_{\sigma}},
\end{equation}
which can be identified as Fano parameter. The real and imaginary
parts of the free retarded conduction electron Green function are:
$ReG_{0,\sigma}(\omega)=P\int
d\epsilon\rho_{w0,\sigma}(\epsilon)/(\omega-\epsilon)$ and
$ImG_{0,\sigma}(\omega)=-\pi\rho_{w0,\sigma}(\omega)$, and
parameters $\Delta_{\sigma}=t_{\gamma}^2ReG_{0,\sigma}$ and the
$\Gamma_{\gamma,\sigma}=\pi t_{\gamma}^2
\rho_{w0,\sigma}(\omega)$. In the non-interacting case when the QD
is represented by a single-particle level,
$G_{QD\gamma,\sigma}=[\omega-\epsilon_{\gamma,\sigma}-\Delta_{\sigma}+i\Gamma_{\gamma,\sigma}]^{-1}$
, and making substitution
$\epsilon=(\omega-\epsilon_{\gamma,\sigma}-\Delta_{\sigma})/\Gamma_{\gamma,\sigma}$
we get the Fano well-known formula from Eq.(~\ref{ro1qd}):
$\rho_{w,\sigma}(\omega)=\rho_{w0,\sigma}(\omega)[(\epsilon+q_{\sigma})^2/(\epsilon^2+1)]$.

For the interacting quantum dot, the Green function in expression
for T-matrix should contain appropriate information on
correlations. Within IPS it has the form:
\begin{equation}
G_{QD1,\sigma}^{INT}(\omega)=\lbrack\omega-\epsilon_{1,\sigma}-\Delta_{\sigma}-
n_{1,-\sigma}U-\Sigma_{1,\sigma}^{IPS}
 +i\Gamma_{1,\sigma}\rbrack^{-1}.
\end{equation}
Thus, the bare quantum dot level is additionally renormalized and
gets the width due to selfenergy $\Sigma^{IPS}$ by
electron-electron interactions and the Kondo peak is generated by
specific $\omega$-dependence of $\Sigma^{IPS}$.

Numerical calculations of the selfenergy within IPS have been
performed under assumption of smooth and weakly $\omega$-dependent
density of states in the wire near the Fermi level. This is a
requirement for stability of numerical procedures when calculating
selfenergy $\Sigma^{IPS}$. Then, the integral defining
$ReG_{0,\sigma}(\omega)=P\int\limits_{-D}^Dd\epsilon\rho_{w0,\sigma}(\epsilon)/(\omega-\epsilon)
\cong\rho_{w0,\sigma}ln\vert(D+\omega)/(D-\omega) \vert\to 0$ for
halfwidth of spectral density of the wire $D\gg\omega$ implying
$q_{\sigma}=0$ (see Eq.(\ref{Fanoq})).

Utilizing Eqs. (\ref{ro1qd}) and (\ref{for4}), at $T=0$ (and for
$q_{\sigma}=0$) the formula for conductance can be written:
\begin{equation}
\label{G1QDwirU0}
 \mathcal{G}=\frac{2\pi e^{2}}{h}\sum_{\sigma}\Gamma_{\sigma}(0)\rho_{w0,\sigma}(0)
\left[1-\frac{\Gamma_{1max,\sigma}^2}{(\epsilon_{\sigma})^2+\Gamma_{1max,\sigma}^2}\right],
\end{equation}
where for non-interacting case
$\epsilon_{\sigma}=\epsilon_{1,\sigma}$ is the bare QD level, and
in presence of interactions
$\epsilon_{\sigma}=\epsilon_{eff1,\sigma}\equiv\epsilon_{1,\sigma}+n_{1,-\sigma}U+Re\Sigma_{1}^{IPS}(0)$
is the renormalized level. In this particular case, when
$Im\Sigma^{IPS}(\omega=0)=$0 as for Fermi liquid, the conductance
for non-interacting and interacting cases can be expressed by the
same relation. It is a consequence of the fact, that Hamiltonian
Eq.(\ref{for1}) in this case can be formally written as for
non-interacting resonant level model, with bare QD levels
$\epsilon_{\gamma,\sigma}$ replaced by renormalized ones
$\epsilon_{eff\gamma,\sigma}$.
\begin{figure}
\begin{center}
\includegraphics*[width=8 cm]{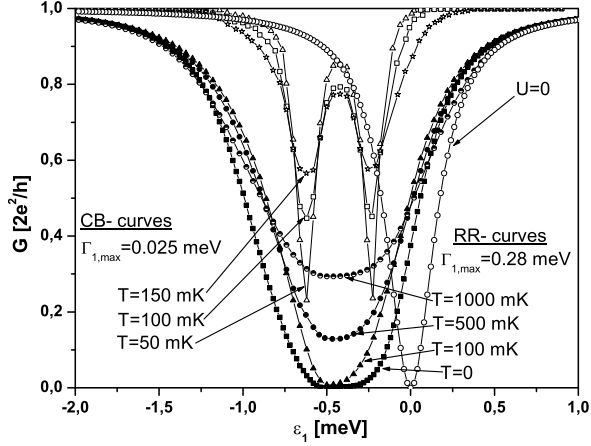}
\end{center}
 \caption{Conductance through the wire with one QD side attached to it vs.
 QD spatial level position. Two regimes are presented:
 resonant regime (RR) and Coulomb blockade (CB) regime.
 Calculations are made for Coulomb repulsion $U=$1meV. Open circles - $U=$0 case.}
 \label{fig1}
\end{figure}
 Conductance vs. QD level position is plotted in Fig.(\ref{fig1}). Two
types of curves are shown: in resonant regime (RR), with one broad
antiresonance, and in Coulomb blockade (CB) regime with two
antiresonances. In the resonant regime, for $T=0$, the total
extinction of conductance is observed for $\epsilon_{\gamma}<0$
and $\epsilon_{\gamma}+U>0$, i.e. when the QD level is singly
occupied. In this case the destructive interference has its full
strength due to Kondo resonance. For $\epsilon_{\gamma}>0$ (empty
level) or $\epsilon_{\gamma}+U<0$ (doubly occupied level) the dot
weakly disturbs transport through the wire and conductance
approaches unitary limit. For this configuration  Kondo
temperature $T_{K}\sim$ 700 mK for
$\epsilon_{\gamma,\sigma}=-U/2$, as obtained from the full width
of the QD density of states at half maximum. When temperature
increases, the Kondo effect is gradually diminished and the
scattering of the electronic waves by many-body QD's Kondo level
becomes less effective. By a decrease of QD coupling to the wire
the system is driven into the Coulomb blockade regime. By this
operation the Fano resonance caused by effective Kondo peak in the
vicinity of Fermi surface is depressed. Instead, two Fano
resonances appear due to scattering on two single particle levels
with the separation of the order of $U$. Although the Kondo
temperature of the QD has been decreased by lowering hopping
$t_{1}$, the electron correlations still exist in the system. When
the temperature is increased the CB peaks get sharper due to
further diminishing of strong electron correlations. For
comparison, the conductance for $U=0$ has been calculated. In this
case the total suppression of conductance takes place when the
bare dot level crosses Fermi level.

How Fano resonances appear conductance in Fig. (\ref{fig1}) can be
better understood in a simplified picture. At temperatures much
lower than Kondo temperature, $T\ll T_{K}$, the Green function of
interacting quantum dot can be approximated as:
\begin{eqnarray}
\nonumber
 G_{QD1,\sigma}^{INT}(\omega+i\delta)\cong\frac{Z_{1}}{\omega-\epsilon_{1,\sigma}-\Delta_{\sigma}+
 i\Gamma_{1,\sigma}}\\\label{analinQD}
 +\frac{Z_{U}}{\omega-\epsilon_{1,\sigma}-\Delta_{\sigma}-U+i\Gamma_{1,\sigma}}
 +\frac{Z_{K}}{\omega-\epsilon_{K}+iT_{K}}
\end{eqnarray}
The first and second terms simulate charge peaks in the spectral
density and the third term represents many-body Kondo peak of the
width proportional to the Kondo temperature $T_{K}$ and position
$\epsilon_{K}$ in the vicinity of Fermi surface. $Z_{1}$, $Z_{U}$
and $Z_{K}$ are appropriate strengths of the poles:
$Z_{1}+Z_{U}+Z_{K} \le 1$. In the Kondo limit the Kondo peak can
be regarded as a potential scatterer  \cite{hewson} and $Z_{K}\sim
\pi T_{K}/\Gamma$ \cite{bick}. Utilizing Eqs.(\ref{analinQD}) and
(\ref{ro1qd}) the spectral density of conduction electron
scattered by QD is written:
\begin{eqnarray}
\label{rofl}
\rho_{w,\sigma}(\omega)=\rho_{w0,\sigma}(\omega)
\left(1+Z_{1}
\frac{q_{\sigma}^2-2q_{\sigma}x_{1}-1}{x_{1}^2+1}+\pi
\frac{q_{\sigma}^2-2q_{\sigma}x_{K}-1}{x_{K}^2+1} \right),
\end{eqnarray}
where following substitutions were made:
$x_{1}=\frac{\omega-\epsilon_{1,\sigma}-\Delta_{\sigma}}{\Gamma_{1,\sigma}}$
and $x_{K}=\frac{\omega-\epsilon_{K}}{T_{K}}$. The second and
third terms produce the shape of Fano resonance. When temperature
is sufficiently low and coupling to the wire is large, electron
interactions prevail in the dot. Then the third term in
Eq.(\ref{rofl}) dominates and the Fano resonance in the
conductance vs. position of the QD spatial level dependence is due
to many-body effects. The maximum of Kondo resonance (which is
transformed into Fano resonance in conductance for present
geometry) takes place for $\epsilon_{1}=-U/2$, i.e. for symmetric
Anderson model. The influence of the second term describing
"classical" Fano effect is practically obscured by the Kondo
effect. It becomes to be visible in the Coulomb Blockade regime,
when Kondo temperature of the system is decreased by lowering the
hybridization strength $\Gamma_{\gamma}$ of QD with the wire.
Thus, the Kondo temperature is the energy scale which governs the
transition between "classical" and "many-body" Fano resonances.

Phase shift acquired due to scattering on the localized level can
be derived when the form of T-matrix is known \cite{hewson}:
$\eta(\omega)=argT(\omega+i\delta)$. For non-interacting
electrons, at the Fermi level $\omega=0$, it has the form
$\eta(\omega=0)=arctan(\frac{\Gamma_{1,\sigma}}{\epsilon_{1,\sigma}})$,
where $\epsilon_{1,\sigma}$ is the bare QD level. Thus, the phase
is changed by $\pi/2$ when QD level $\epsilon_{1,\sigma}$ crosses
the Fermi level. For interacting case $\eta(\omega=0)\cong
arctan(\frac{T_{K}}{\epsilon_{K}})$, because the resonant level is
of many-body nature. The position of the Kondo peak is
$\epsilon_{K}\simeq 0$, which gives $\eta(\omega=0)\cong \pi/2$.

\section{Two quantum dots side attached to the wire}
The formula for the Green function of conduction electron
scattered by two quantum dots was obtained from Eq.(\ref{dys1}),
for the selfenergy  given by Eq.(\ref{sigtot}) and total T-matrix
given by Eq.(\ref{totalT}). The spectral density has the form of:
\begin{eqnarray}
 \nonumber
\rho_{w,\sigma}(\omega)=\rho_{w0,\sigma}(\omega)\times\\
\label{spect2qd}\left[
1-\frac{[\Gamma_{1,\sigma}(\omega-\epsilon_{2,\sigma})+\Gamma_{2,\sigma}(\omega-\epsilon_{1,\sigma})]^2}
{[(\omega-\epsilon_{1,\sigma})(\omega-\epsilon_{2,\sigma})]^2+[\Gamma_{1,\sigma}(\omega-\epsilon_{2,\sigma})
+\Gamma_{2,\sigma}(\omega-\epsilon_{1,\sigma})]^2}\right],
\end{eqnarray}
and conductance at T=0 can be expressed as follows:
\begin{equation}
\label{Q2qdU0}
 \mathcal{G}=\frac{2\pi e^{2}}{h}\sum_{\sigma}\Gamma_{\sigma}(0)\rho_{w0,\sigma}(0)
\left[1-\frac{[\epsilon_{1,\sigma}\Gamma_{2max,\sigma}+\epsilon_{2,\sigma}\Gamma_{1max,\sigma}]^2}
{(\epsilon_{1,\sigma}\epsilon_{2,\sigma})^2+[\epsilon_{1,\sigma}\Gamma_{2max,\sigma}+\epsilon_{2,\sigma}\Gamma_{1max,\sigma}]^2}\right].
\end{equation}

Thus, $\mathcal{G}=0$ when $\epsilon_{1}=0$ or $\epsilon_{2}=0$ is
tuned to the Fermi surface in the wire. On the other hand, for
$\epsilon_{\gamma}=(-\Gamma_{\gamma}/\Gamma_{\gamma'})\epsilon_{\gamma'}$
conductance reaches unitary limit at $T=0$. For such condition the
total phase shift
$\eta(\omega)=-arctan(\sum_{\gamma=1,2}\frac{\Gamma_{\gamma}}{\omega-\epsilon_{\gamma}})$
is equal to zero and the electronic wave propagates through the
wire without scattering. Thus, by setting a proper configuration
of the QD's levels by gate voltages, the conductance through
device can be set to unity or to zero.

When the level of one QD, say $\epsilon_{2}$, is kept fixed, while
$\epsilon_{1}$ is shifted by the gate voltage, then the Fano-like
resonances appear in the conductance. The QD level $\epsilon_{2}$
acts as additional background channel and the shape of Fano
resonance depends on its position with respect to the Fermi level.
Indeed, equation (\ref{Q2qdU0}) can be approximated by the
expression:
\begin{eqnarray}
 \mathcal{G}\simeq \frac{2\pi
e^{2}}{h}\sum_{\sigma}\Gamma_{\sigma}(0)\rho_{w0,\sigma}(0)
\label{fanoapp}\left[1-\frac{\Gamma_{max2,\sigma}^2}{\epsilon_{2,\sigma}^2+\Gamma_{max2,\sigma}^2}
\frac{(\epsilon_{\sigma}+q_{\sigma})^2}{\epsilon_{\sigma}^2+1}\right],
\end{eqnarray}
where following substitutions have been made:
$\epsilon_{\sigma}=\epsilon_{1,\sigma}/\Gamma_{max1,\sigma}$ and
Fano parameter
$q_{\sigma}=\epsilon_{2,\sigma}/\Gamma_{max2,\sigma}$.

The non-conservation of the orbital quantum number while hopping
of electrons between quantum dots and wire is crucial for
obtaining Fano-like shaped conductance. In the opposite case no
information could flow from one quantum dot to the other and the
dots would act as separate entities.

To study this problem in detail we considered the Hamiltonian,
Eq.(\ref{for1}), without Coulomb interactions which is quadratic
and can be solved exactly. EOM method gives the expression for
Fourier transformed retarded Green function
$G_{\gamma,\sigma}(t-t')=\langle\langle
d_{\gamma,\sigma}(t);d_{\gamma,\sigma}^{+}(t')\rangle\rangle$ of
$\gamma$-QD level:
\begin{equation}
\label{app1}
G_{\gamma,\sigma}(\omega)=\frac{1}{\omega-\epsilon_{\gamma,\sigma}+
i\Gamma_{\gamma,\sigma}+\frac{\Gamma_{\gamma,\sigma}\Gamma_{\gamma',\sigma}}{\omega-
\epsilon_{\gamma',\sigma}+i\Gamma_{\gamma',\sigma}}}
\end{equation}
Thus, apart from the broadening caused by the hopping of electrons
between $\gamma-$ QD and wire, the $\gamma$-level is additionally
shifted and broadened due to indirect interaction with
$\gamma'$-QD level.

Fourier transformed Green function
$G_{\gamma\gamma'\sigma}(t-t')=\langle\langle
d_{\gamma,\sigma}(t);d_{\gamma',\sigma}^{+}(t') \rangle\rangle$
describing virtual hopping between $\gamma$ and $\gamma'$ levels
has the form:
\begin{equation}
\label{app2}
G_{\gamma\gamma',\sigma}(\omega)=\frac{-i\Gamma_{\gamma\gamma',\sigma}}
{(\omega-\epsilon_{\gamma,\sigma})(\omega-\epsilon_{\gamma',\sigma})+i[\Gamma_{\gamma,\sigma}(\omega-
\epsilon_{\gamma',\sigma})+\Gamma_{\gamma',\sigma}(\omega-
\epsilon_{\gamma,\sigma})]},
\end{equation}
 where $\Gamma_{\gamma\gamma',\sigma}=\pi
t_{\gamma}^{*}t_{\gamma'}\rho_{w0,\sigma}$.
 Utilizing Eqs.(\ref{app1}-\ref{app2}) and assuming $t_{\gamma}$ ($\gamma=1,2$) real, the expression
for the total T-matrix due to both quantum dots can be rewritten
as:
\begin{eqnarray}
 \nonumber
T_{\sigma}=t_{1}^{2}G_{QD1,\sigma}+t_{2}^{2}G_{QD2,\sigma}+t_{1}t_{2}G_{12,\sigma}+t_{2}t_{1}G_{21,\sigma}\\
\label{t2qd}=T_{1,\sigma}+T_{2,\sigma}+2T_{12,\sigma},
\end{eqnarray}
where
$T_{\gamma\gamma'}=t_{\gamma}t_{\gamma'}G_{\gamma\gamma',\sigma}$
and $T_{\gamma\gamma',\sigma}=T_{\gamma'\gamma.\sigma}$. The
spectral density, Eq.(\ref{spect2qd}), can now be written in a
more transparent form:
\begin{eqnarray}
\nonumber
 \rho_{w,\sigma}(\omega)=\rho_{w0,\sigma}(\omega)\left[
1+\Gamma_{1,\sigma}ImG_{QD1}(\omega)\right.\\
\label{rotrans}+\left.\Gamma_{2,\sigma}ImG_{QD2}(\omega)+2\Gamma_{12}ImG_{12}\right].
\end{eqnarray}
Eq.(\ref{t2qd}) consists of terms describing scattering on each
$\gamma$ quantum dot separately (but with renormalization due to
indirect interaction with $\gamma'$ quantum dot), and the term of
multiple scattering. When the $\gamma$- quantum dot is gradually
decoupled from the system (by appropriate decrease of coupling
$t_{\gamma}$, or by shifting $\epsilon_{\gamma}$ far from the
Fermi level) both the $\gamma$-term and last multiple scattering
term vanish and Eq.(\ref{rotrans}) coincides with Eq.(\ref{ro1qd})
(with $q_{\sigma}=0$) for a single quantum dot.
 When both
$\gamma=1,2$ levels are close to the Fermi surface, the electronic
wave traveling through the wire scatters resonantly in both
quantum dots. Classically, this process can be seen as a multiple
bouncing of the ball between two walls. This process is summed up
to infinity by the Dyson equation Eq.(\ref{2qddys}). The maximum
of the "bouncing effect" on the Fermi surface (for $\omega=0$) due
to the last term in Eq.(\ref{rotrans}) takes place for
$\epsilon_{\gamma}=\frac{\Gamma_{\gamma}}{(\epsilon_{\gamma'}^2+\Gamma_{\gamma'}^2)^{1/2}}(-\epsilon_{\gamma'})$,
i.e. when the levels lie below and above Fermi surface.

When strong electron correlations within the dots are included,
the explicit expression for conduction electron Green function
Eq.(\ref{dys1}) has complicated form, but it simplifies
considerably at $T=0$ (see the comment below
Eq.(\ref{G1QDwirU0})). The conductance can then be written in
exactly the same form as for non-interacting case
(Eq.(\ref{spect2qd}, \ref{Q2qdU0} and \ref{fanoapp})), where
single-particle $\epsilon_{\gamma,\sigma}$ levels have to be
replaced by renormalized ones:
$\epsilon_{eff\gamma,\sigma}=\epsilon_{\gamma,\sigma}+n_{\gamma,-\sigma}U+Re\Sigma_{\gamma}^{IPS}(\omega=0)$
for $\gamma=1,2$. The discussion of Eqs.(\ref{t2qd} and
\ref{rotrans}) also holds with the above replacement. In this
special case when $Im\Sigma^{IPS}=0$ and the bare QD levels are
renormalized by static field
$H_{eff}=n_{\gamma,-\sigma}U+Re\Sigma_{\gamma}^{IPS}(\omega=0)$,
all the dynamical many-body effects are erased. In general, apart
from renormalized by interactions single-particle levels, Kondo
resonances are present in the vicinity of Fermi surface which have
dominant influence on conductance.
\begin{figure}
\begin{center}
\includegraphics*[width=8 cm]{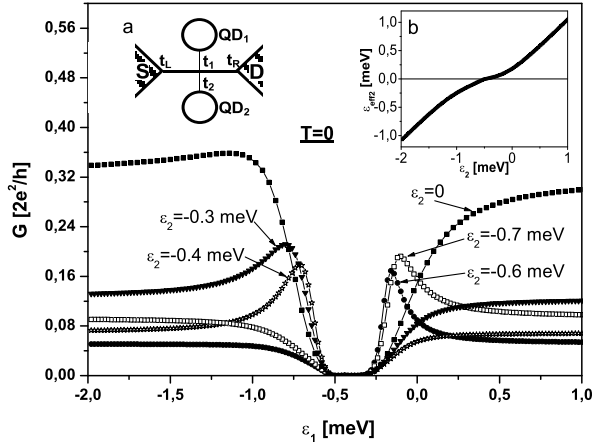}
\end{center}
 \caption{Conductance through a wire with two QDs side-attached to it vs.
 spatial level position $\epsilon_{1}$ in $QD_{1}$ for various values of level $\epsilon_{2}$ in $QD_{2}$.
 Inset $a$ - scheme of the device, inset $b$ - dependence of renormalized level $\epsilon_{eff2}$ level vs. bare
 $\epsilon_{2}$ level position.}
 \label{fig2}
\end{figure}

Conductance dependence on the $\epsilon_{1}$ QD level position for
various values of level $\epsilon_{2}$ is plotted in
Fig.(\ref{fig2}) at $T=0$ and in presence of correlations. The
change of the Fano resonance shape can be observed as
$\epsilon_{2}$ takes different values. A mirror reflection of the
curve takes place when $\epsilon_{eff2,\sigma}$ changes sign
changing $q_{\sigma}=\epsilon_{eff2,\sigma}/\Gamma_{2,\sigma}$
parameter. The $\epsilon_{eff2}$ dependence on bare $\epsilon_{2}$
is shown in the $b$-inset.

\begin{figure}
\begin{center}
\includegraphics*[width=8 cm]{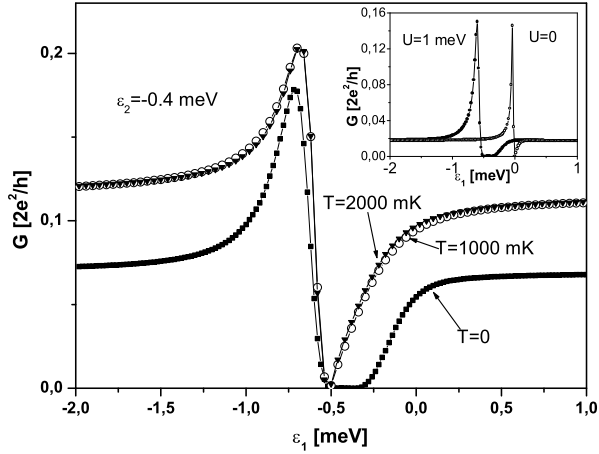}
\end{center}
 \caption{Conductance through a wire with two QDs side-attached to it vs. spatial level
 position $\epsilon_{1}$ in $QD_{1}$ for a fixed value of level $\epsilon_{2}$=-0.4 meV in $QD_{2}$ and $U=$1 meV within the dots.
 Apart from T=0 curve, two curves are plotted for temperatures much higher than the Kondo temperature. Inset: conductance curves
 for interacting and non-interacting cases calculated for the same Fano parameter $q$.}
 \label{fig3}
\end{figure}

 To compare the conductance curves
calculated with and without interactions we show two of them in
the inset of Fig.(\ref{fig3}). The curve for $U=0$ has been
plotted for $\epsilon_{2}$ equal to corresponding
$\epsilon_{eff2}$ found selfconsistently in interacting case and
the same $\Gamma_{2,\sigma}$ to have the same Fano parameter in
both cases. Two substantial differences are seen between the
curves: i) the curves are shifted; this is caused by
renormalization of bare $\epsilon_{1}$ level by interactions, ii)
an additional plateau emerges for $U\neq 0$ which is caused by
Kondo resonant scattering when the $\epsilon_{1}$ level is
occupied by one electron.

The $q=0$ case deserves a comment. In this case $\epsilon_{eff2}$
coincides with Fermi level, which corresponds to the bare
$\epsilon_{2,\sigma}=-U/2$ i.e. maximum Kondo resonance in
$QD_{2}$. For such level arrangement conductance $\mathcal{G}=0$
for any value of shifted $\epsilon_{1}$ because of full
destructive interference caused by $QD_{2}$. This is a different
situation comparing to classical Fano expression which for $q=0$
describes a symmetric dip. This shows a resonant nature of the
background channel provided by the second quantum dot.

In Fig.(\ref{fig3}) we show conductance curves for fixed $QD_{2}$
level position $\epsilon_{2}=$-0.4 meV at $T=0$ and at
temperatures much higher than Kondo temperature. At elevated
temperatures the plateau in conductance  due to Kondo effect on
$QD_{1}$ disappears, but the shape of the Fano resonance
practically does not change. It leads to the conclusion that the
shape of the Fano resonance is determined by the renormalized
single particle level $\epsilon_{eff2}$ and not by the many-body
Kondo peak caused by resonant scattering on $QD_{2}$.

\section{Acknowledgments} This work
was supported by the Polish State Committee for Scientific
Research (KBN) under Grant no. PBZ/KBN/044/P03/2001.

\end{document}